\documentclass[reprint,showpacs,amsmath,amssymb,pra]{revtex4-1}
\usepackage{graphicx}
\usepackage{dcolumn}
\usepackage{bm}
\usepackage{color}

\begin{document}

\title{$\mathcal{PT}$-symmetry breaking in multilayers with resonant loss and gain locks light propagation direction}

\author{Denis~V.~Novitsky$^{1,2,3}$}
\email{dvnovitsky@gmail.com}
\author{Alina~Karabchevsky$^{4,5,6}$}
\author{Andrei~V.~Lavrinenko$^7$}
\author{Alexander~S.~Shalin$^2$}
\author{Andrey~V.~Novitsky$^{7,8}$}

\affiliation{$^1$B.~I.~Stepanov Institute of Physics, National
Academy of Sciences of Belarus, Nezavisimosti Avenue 68, 220072
Minsk, Belarus \\ $^2$ITMO University, Kronversky Prospekt 49,
197101 St. Petersburg, Russia \\ $^3$Saint Petersburg
Electrotechnical University ``LETI'', 5 Prof. Popova Str., 197376
St. Petersburg, Russia \\ $^4$ Electrooptical Engineering Unit,
Ben-Gurion University of the Negev, David Ben-Gurion Blvd, P.O.B.
653, 8410501 Beer-Sheva, Israel \\ $^5$ Ilse Katz Institute for
Nanoscale Science $\&$ Technology, Ben-Gurion University of the
Negev, Beer-Sheva, 8410501, Israel \\ $^6$ Center for Quantum
Information Science and Technology, Ben-Gurion University of the
Negev, Beer-Sheva, 8410501, Israel \\ $^7$ DTU Fotonik, Technical
University of Denmark, {\O}rsteds Plads 343, DK-2800 Kongens Lyngby,
Denmark \\ $^8$ Department of Theoretical Physics and Astrophysics,
Belarusian State University, Nezavisimosti Avenue 4, 220030 Minsk,
Belarus}

\begin{abstract}
Using the Maxwell-Bloch equations for resonantly absorbing and
amplifying media, we study the temporal dynamics of light
propagation through the $\mathcal{PT}$-symmetric structures with
alternating loss and gain layers. This approach allows us to
precisely describe the response of the structure near the
exceptional points of $\mathcal{PT}$-symmetry breaking phase
transition and, in particular, take into account the nonlinear
effect of loss and gain saturation in the $\mathcal{PT}$-symmetry
broken state. We reveal that in this latter state the multilayer
system possesses a lasing-like behavior releasing the pumped energy
in the form of powerful pulses. We predict locking of pulse
direction due to the $\mathcal{PT}$-symmetry breaking, as well as
saturation-induced irreversibility of phase transition and
nonreciprocal transmission.
\end{abstract}

\maketitle

\section{Introduction}

$\mathcal{PT}$-symmetric structures and $\mathcal{PT}$-symmetry
breaking seem to be one of the most intensively studied fields in
optics and photonics of active systems
\cite{Zyablovsky2014,Feng2017,El-Ganainy2018}. It dates back to the
seminal works of Bender and Boettcher \cite{Bender1998,Bender2007},
who discovered the real-valued spectra of non-Hermitian Hamiltonians
in quantum mechanics provided these Hamiltonians are parity-time
($\mathcal{PT}$) symmetric, i.e., invariant with respect to
simultaneous parity change and time reversal. Such ideas can be
straightforwardly transferred to optics using the spatial ordering
of passive and active components, thus introducing the concept of an
optical $\mathcal{PT}$-symmetric system.

The simplest optical $\mathcal{PT}$-symmetric structure can be
realized by means of one-dimensional multilayers --- similar to
photonic crystals --- with proper spatial variation of the complex
permittivity as $\varepsilon(z) = \varepsilon^\ast(-z)$ (here
asterisk stands for complex conjugation). This condition implies
that the real part of the permittivity or refractive index is an
even function of the coordinate, whereas the imaginary part is an
odd function. Change of the sign of the imaginary part of the
permittivity ${\rm Im} \varepsilon(z) = -{\rm Im} \varepsilon(-z)$
apparently requires amplifying materials. It can be, for instance,
an alternation of loss and gain in periodic multilayers. Two
principal schemes, longitudinal and transverse ones, are usually
employed. In the former system, light propagates directly through
the multilayer. In the latter scheme light propagates
perpendicularly to the permittivity distribution, along the layers
boundaries, being analogous to a system of interconnected waveguides
(an optical grating). First theoretical
\cite{El-Ganainy2007,Makris2008} and experimental \cite{Ruter2010}
results on optical $\mathcal{PT}$ symmetry were reported exactly for
the coupled waveguides.

$\mathcal{PT}$ symmetry allows new ways for controlling radiation
fluxes both in optics and plasmonics \cite{Yang2015}. A number of
effects can be highlighted as a fingerprint of the $\mathcal{PT}$
symmetry: nonreciprocity of light transmission and beam power
oscillations \cite{Makris2008}, anisotropic transmission resonances
\cite{Ge2012}, unidirectional ``invisibility'' \cite{Lin2011},
negative refraction, and focusing of light \cite{Fleury2014}.
$\mathcal{PT}$ symmetry governs light localization in disordered
structures \cite{Kartashov2016}. The usage of gain media raises
questions concerning available nonlinear phenomena, such as optical
switching and generation of new types of solitons
\cite{Suchkov2016,Konotop2016}. Apodization of the refractive index
spatial profiles can be used to facilitate switching conditions in
$\mathcal{PT}$-symmetric Bragg gratings \cite{Lupu2016}. Some
nonlinear effects connected to $\mathcal{PT}$ symmetry have been
recently observed in experiments with coherent atomic gases
\cite{Hang2017}. More prospects are opened by the fact that
$\mathcal{PT}$-symmetric optical gratings are capable of supporting
topologically protected bound states \cite{Weimann2017}.

It should be noted that the effects mentioned above may be
observable in loss-gain structures lacking the
$\mathcal{PT}$-symmetry. A typical example is the reflectionless
transmission and unidirectional ``invisibility'' for both normal
\cite{Shen2014,Ramirez2017} and oblique \cite{Novitsky2017}
incidence. However, the symmetry breaking at the exceptional points
belongs exclusively to the $\mathcal{PT}$ symmetry domain.

There is a number of phenomena associated with violation of the
$\mathcal{PT}$ symmetry. First, a sharp change in polarization
response of the system is possible at the exceptional points
\cite{Lawrence2014}. Later an omnipolarizer was designed for
converting any light polarization into a given one
\cite{Hassan2017}. Second, enhanced sensitivity of such kind of
systems to external perturbations near exceptional points provides a
new approach for sensing \cite{Chen2017,Hodaei2017}. Third,
$\mathcal{PT}$-symmetry breaking plays an important role in laser
physics offering new types of lasers
\cite{Feng2014,Hodaei2014,Gu2016} and anti-lasers \cite{Wong2016}
based on the effect of coherent perfect absorption
\cite{Chong2010,Longhi2010}. Finally, the possibility of light
stopping at the exceptional point was recently reported
\cite{Goldzak2018}.

In this paper, we study the phenomenon of $\mathcal{PT}$-symmetry
breaking in one-dimensional multilayers composed of resonantly
absorbing and amplifying media. We describe propagation of light in
the time domain using the Maxwell-Bloch equations and taking into
account loss and gain saturation. The influence of the saturable
nonlinearity on nonreciprocity and bistability of
$\mathcal{PT}$-symmetric structures was previously reported both for
transverse \cite{Ramezani2010} and longitudinal geometries
\cite{Phang2014,Liu2014,Barton2017,Witonski2017}. However, those
investigations introduced the saturation phenomenologically via the
permittivity. Our approach is based on self-consistent description
of temporal dynamics for both light field and medium loss/gain
exhibiting a more realistic treatment of $\mathcal{PT}$-symmetric
optical systems. Therefore, we have a deep insight into
peculiarities of the $\mathcal{PT}$-symmetry breaking and related
effects. In particular, we study the lasing-like regime in the
$\mathcal{PT}$-symmetry broken state, where the saturation effects
lead to the unique phase transition in the parameter space and
nonreciprocal transmission of generated pulses. Pulse-direction
locking by the $\mathcal{PT}$-symmetry breaking in the lasing-like
regime is related to the strong light confinement and resembles
light polarization locking to its propagation direction in quantum
optics \cite{Lodahl17}.

The paper is organized as follows. Section \ref{eqpars} is devoted
to the description of the theoretical model of the loss-gain multilayer
and parameters used in calculations. We discuss light behavior in
the $\mathcal{PT}$-symmetric phase in Sec. \ref{stat} by comparing
numerical solution of the Maxwell-Bloch equations and stationary
transfer-matrix calculations. In Sec. \ref{trans}, the
$\mathcal{PT}$-symmetry breaking as a phase transition to the
lasing-like regime is studied with emphasis on temporal dynamics of the
light propagating through the multilayers. Section \ref{concl}
summarizes the article.

\section{\label{eqpars} Resonant loss and gain media}

\begin{figure}[t!]
\includegraphics[scale=0.3, clip=]{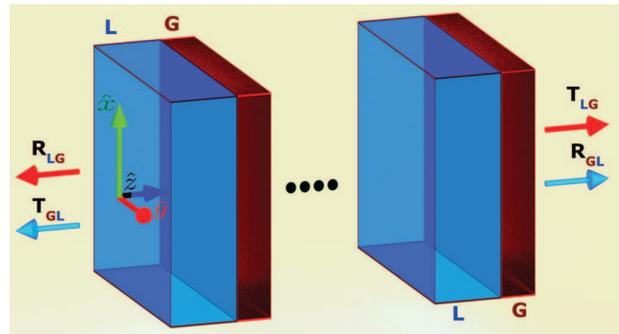}
\caption{\label{fig1} $N$-periods multilayer system of alternating
slabs with loss (L) and gain (G). Transmissions and reflections and
corresponding propagation directions are indicated for the waves
incident from left-hand ($T_{LG}$ and $R_{LG}$) and right-hand
($T_{GL}$ and $R_{GL}$) sides. }
\end{figure}

A periodic planar structure composed of $2N$ alternating loss and
gain layers shown in Fig.~\ref{fig1} is illuminated by monochromatic
light of angular frequency $\omega$ at normal incidence. In this
study, both loss and gain are described in the similar manner, using
the model of a homogeneously-broadened two-level medium. Excluding
the rapidly varying factors $\exp(-i\omega t)$ in polarization,
population difference, and electric field, we write the
Maxwell-Bloch equations \cite{Novitsky2011} for slowly varying
amplitudes of these quantities as
\begin{eqnarray}
\frac{d\rho}{d\tau}&=& i l \Omega w + i \rho \delta - \gamma_2 \rho, \label{dPdtau} \\
\frac{dw}{d\tau}&=&2 i (l^* \Omega^* \rho - \rho^* l \Omega) -
\gamma_1 (w-w_{eq}),
\label{dNdtau} \\
\frac{\partial^2 \Omega}{\partial \xi^2}&-& n_d^2 \frac{\partial^2
\Omega}{\partial \tau^2}+2 i \frac{\partial \Omega}{\partial \xi}+2
i n_d^2 \frac{\partial \Omega}{\partial
\tau} + (n_d^2-1) \Omega \nonumber \\
&&=3 \alpha l \left(\frac{\partial^2 \rho}{\partial \tau^2}-2 i
\frac{\partial \rho}{\partial \tau}-\rho\right), \label{Maxdl}
\end{eqnarray}
where $\tau=\omega t$ and $\xi=kz$ are respectively the
dimensionless time and distance, $\Omega=(\mu/\hbar \omega) A$ is
the normalized Rabi frequency, $A$ is the electric field strength,
$\omega$ is the light circular frequency, $k = \omega/c$ is the
wavenumber in vacuum, $c$ is the speed of light, $\hbar$ is the
reduced Planck constant, and $\mu$ is the dipole moment of the
quantum transition. Rabi frequency is dynamically coupled to the
characteristics of the two-level system -- complex amplitude of
microscopic (atomic) polarization $\rho$ and difference between
populations of ground and excited states $w$. Efficiency of the
light-matter coupling is given by the dimensionless parameter
$\alpha= \omega_L / \omega = 4 \pi \mu^2 C/3 \hbar \omega$, where
$\omega_L$ is the Lorentz frequency and $C$ is the concentration
(density) of active (two-level) atoms. In general, light frequency
$\omega$ is detuned from frequency $\omega_0$ of the atomic
resonance as described by $\delta=(\omega_0-\omega)/\omega$. The
normalized relaxation rates of population $\gamma_1=1/(\omega T_1)$
and polarization $\gamma_2=1/(\omega T_2)$ are expressed by means of
the longitudinal $T_1$ and transverse $T_2$ relaxation times. The
influence of the polarization of the background dielectric having
real-valued refractive index $n_d$ on the embedded active particles
is taken into account by the local-field enhancement factor
$l=(n_d^2+2)/3$ \cite{Crenshaw2008,Bloembergen}.

Equilibrium population difference $w_{eq}$ will allow us to describe
both gain and loss materials with the same Maxwell-Bloch equations
(\ref{dPdtau})-(\ref{Maxdl}). When external pump is absent, the
two-level atoms are in the ground state ($w_{eq}=1$), and the medium
is lossy. In the case of gain, the equilibrium population difference
can be referred to as a pumping parameter. In the fully inverted
medium, all atoms are excited by the pump ($w_{eq}=-1$). For the
saturated medium with both levels populated equally in the
equilibrium, there are no transitions between the levels
($w_{eq}=0$).

In the steady-state approximation, when the amplitudes of population
difference, polarization, and field are time-independent, one can
use the effective permittivity of a two-level medium
\cite{Novitsky2017}
\begin{eqnarray}
\varepsilon_{eff} &=& n_d^2+4 \pi \mu C \rho_{st}/E
= n_d^2+\frac{K(-\delta+i\gamma_2)}{1+|\Omega|^2/\Omega^2_{sat}},
\label{epsTLM}
\end{eqnarray}
where $\Omega_{sat}=\sqrt{\gamma_1 (\gamma_2^2+\delta^2)/4l^2
\gamma_2}$ sets the level of saturation intensity and $K=3 \omega_L
l^2 w_{eq}/[\omega (\gamma_2^2+\delta^2)]$. At the exact resonance
$\delta=0$ and in approximation of low-intensity external radiation
$|\Omega| \ll\Omega_{sat}$, Eq. (\ref{epsTLM}) transforms to
$\varepsilon_{eff} \approx n_d^2+3 i l^2 \omega_L T_2 w_{eq}$. From
this equation it is clear that gain and loss correspond to negative
and positive $w_{eq}$, respectively. In the stationary
approximation, it is straightforward to obtain a
$\mathcal{PT}$-symmetric structure composed of alternating layers
with balanced loss ($\varepsilon_{eff+}$) and gain
($\varepsilon_{eff-}$), where
\begin{eqnarray}
\varepsilon_{eff\pm} \approx n_d^2 \pm 3 i l^2 \omega_L T_2
|w_{eq}|. \label{epsPT}
\end{eqnarray}
$\mathcal{PT}$ symmetry holds true, because the necessary condition
$\varepsilon(z) = \varepsilon^\ast(-z)$ is fulfilled, providing even
(odd) function of $z$ for the real (imaginary) part of the
permittivity. In Supplemental Material \cite{Supp},
$\mathcal{PT}$-symmetry conditions are derived straight from the
Maxwell-Bloch equations. It is shown that the system is
$\mathcal{PT}$-symmetric only in a steady state established after
some transient period.

Identity of the absolute values of the imaginary parts of
permittivities $\varepsilon_{eff+}$ and $\varepsilon_{eff-}$ can be
achieved in different ways. From a practical point of view, it would
be convenient to take unexcited absorbing layers ($w_{eq,L}=1$) and
pump only the amplifying layers to the level $w_{eq,G} = -
\alpha_{L}/\alpha_{G}$, where $\alpha_{L}$ and $\alpha_{G}$ are the
light-matter coupling coefficients for loss and gain layers,
respectively. Tuning of these coefficients can be properly carried
out by affecting the concentration of active particles in both types
of layers. Without imposing any restrictions, it is fair to claim
within this theoretical investigation that the loss and gain layers
have equal concentrations $C$ (hence, equal couplings $\alpha$) and
absolute values of the pumping parameter $|w_{eq}|$. Some additional
data on the variant with completely unexcited absorbing layers are
given in Supplemental Material \cite{Supp}.

Eqs. (\ref{dPdtau})--(\ref{Maxdl}) are solved numerically using the
FDTD approach developed in our previous publication
\cite{Novitsky2009} and recently adapted to study loss-gain
structures \cite{Novitsky2017}. As an initial value of the
population difference, we employ the pumping parameter, i.e.,
$w(t=0)=w_{eq}$. Comparison of the results of numerical simulations
with those of the transfer-matrix method with Eq. (\ref{epsPT}) for
permittivities of loss and gain layers will unveil the limitations
of applicability of the latter approach.

In this paper, we use semiconductor doped with quantum dots as an
active material and assume the condition of exact resonance
$\delta=0$ is valid. It can be characterized by the following
parameters \cite{Palik,Diels}: $n_d=3.4$, $\omega_L=10^{11}$
s$^{-1}$, $T_1=1$ ns, and $T_2=0.5$ ps. Gain coefficient $g=4 \pi
\textrm{Im}(\sqrt{\varepsilon_{eff-}})/\lambda \leq 10^4$
cm$^{-1}$ is estimated according to Eq. (\ref{epsPT}) for $\lambda
\sim 1.5$ $\mu$m and $|w_{eq}| \leq 0.2$ can be realized in
practice \cite{Babicheva12}. The multilayer structure contains
$N=20$ unit cells. Both loss and gain layers have the same thickness
$d=1$ $\mu$m. The pumping scheme similar to that realized by Wong
\textit{et al.} \cite{Wong2016} can be used in our system. It is
also worth noting that the choice of materials is not unique, but
the multilayer parameters and light wavelength may need to be
appropriately adjusted to obtain similar results with different
materials.

\section{\label{stat}Temporal dynamics of light in $\mathcal{PT}$-symmetric phase}

\begin{figure*}[t!]
\includegraphics[scale=1.5, clip=]{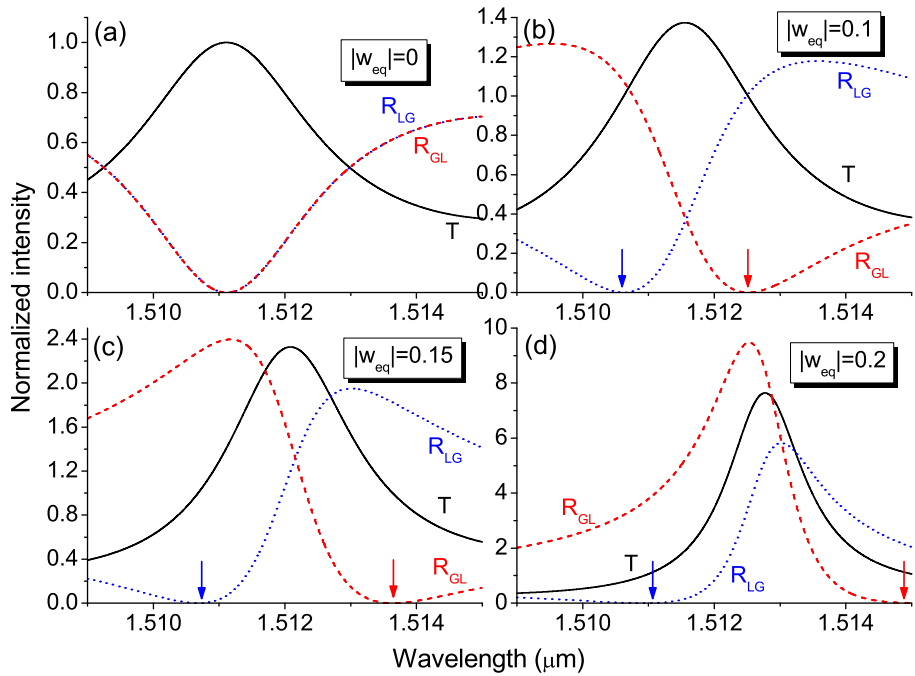}
\centering\caption{\label{fig2} Reflection (R) and transmission (T)
spectra for different levels of pumping: (a) $|w_{eq}|=0$, (b)
$|w_{eq}|=0.1$, (c) $|w_{eq}|=0.15$, and (d) $|w_{eq}|=0.2$. The
parameters of the $\mathcal{PT}$-symmetric structure are given in
the text. Arrows mark anisotropic transmission resonances.}
\end{figure*}

We start our study analysing the stationary characteristics of
one-dimensional $\mathcal{PT}$-symmetric structures. In the
stationary mode, the transfer-matrix method for the wave propagation
through multilayers with permittivities (\ref{epsPT}) is exploited.
Owing to reciprocity of the system, the transmission of oppositely
propagating (forward and backward) waves is the same, but the
reflection is different. We denote these two directions of wave
propagation with subscripts LG and GL (see Fig. \ref{fig1})
originating from the order of layers in the unit cell of the
structure. In the case of $w_{eq} = 0$ (homogeneous dielectric slab
of thickness $2Nd$), the reflections are equal, $R_{LG}=R_{GL}$
[Fig. \ref{fig2}(a)]. Divergence of the curves for $R_{LG}$ and
$R_{GL}$ in Figs. \ref{fig2}(b)-\ref{fig2}(d) indicates existence of
the $\mathcal{PT}$-symmetry in the full accordance with the
properties of transfer matrices of such systems \cite{Ge2012}. In
these figures, one can notice another well-known feature -- the
so-called anisotropic transmission resonance (ATR) \cite{Ge2012}. It
emerges under the following conditions: transmission $T=1$ and one
of the reflections is zero. Thus, the ATR arises for the two
wavelengths marked with arrows in Fig. \ref{fig2} corresponding to
$R_{GL} > R_{LG}=0$ and $R_{LG} > R_{GL}=0$. Transmission exceeds
unity between these points.

\begin{figure}[t!]
\includegraphics[scale=1., clip=]{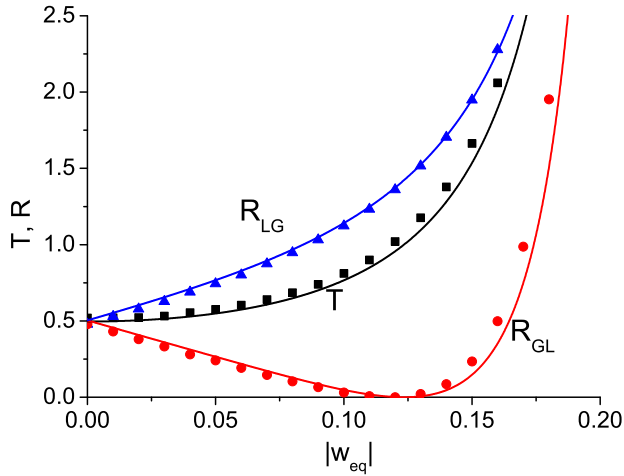}
\centering \caption{\label{fig3} Dependence of stationary levels of
the reflection (R) and transmission (T) on pumping parameter
$|w_{eq}|$ at the wavelength $\lambda=1.513$ $\mu$m. Symbols and
lines correspond to results of numerical and transfer-matrix
calculations, respectively.}
\end{figure}

\begin{figure}[t!]
\includegraphics[scale=1., clip=]{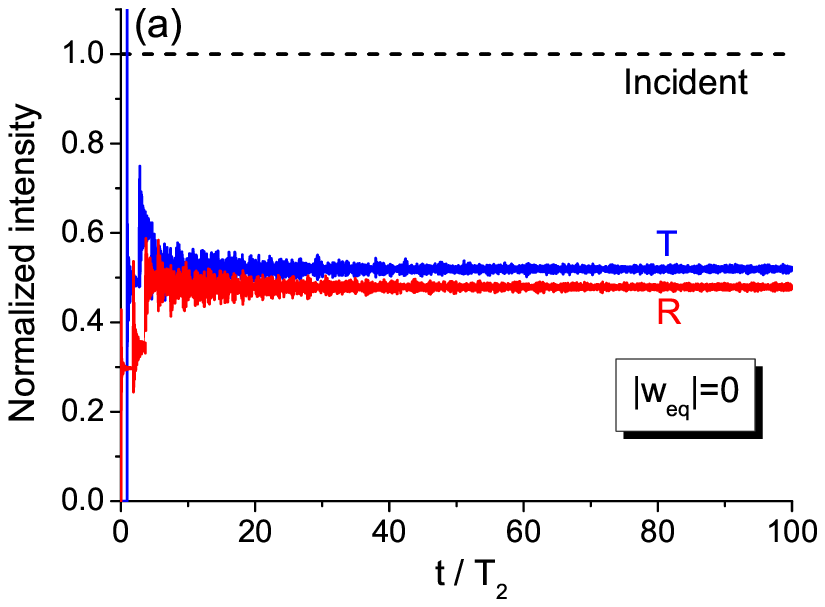}
\includegraphics[scale=1., clip=]{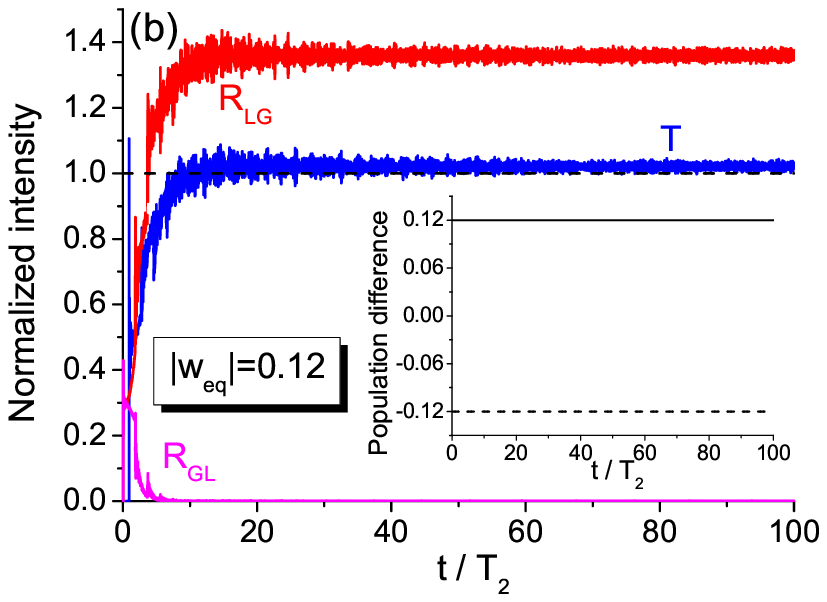}
\centering \caption{\label{fig4} Temporal dynamics of the reflected
(R) and transmitted (T) intensity for the pumping parameters (a)
$|w_{eq}|=0$ and (b) $|w_{eq}|=0.12$. Inset shows dynamics of
population difference at the entrance of the first loss and gain
layers.}
\end{figure}

The increase of the pumping parameter (hence, the absolute value of
the imaginary part of slabs' permittivities) results in the red
shift of the ATR. This means that the ATR can be reached by changing
$|w_{eq}|$ at any fixed wavelength near the transmission peak. In
order to examine this prediction of the transfer-matrix method, we
perform numerical simulations of the Maxwell-Bloch equations
(\ref{dPdtau})-(\ref{Maxdl}) for a monochromatic wave
($\lambda=1.513$ $\mu$m) propagating through the
$\mathcal{PT}$-symmetric multilayer with different pumping
parameters. To keep the correspondence with the transfer-matrix
approach, the saturation should be neglected. Here it is realized
for as low incident wave amplitude as $\Omega_0=10^{-5} \gamma_2 \ll
\Omega_{sat}$. The results of numerical calculations shown with
symbols in Fig. \ref{fig3} agree well with those of matrix method
(lines in Fig. \ref{fig3}). The ATR at $|w_{eq}| \sim 0.125$ is also
confirmed in the FDTD calculations. Temporal dynamics shown in Fig.
\ref{fig4} demonstrates the transient process to the steady-state
formation in the dielectric slab [in the absence of loss and gain,
panel (a)] and $\mathcal{PT}$-symmetric structure in conditions of
the ATR [panel (b)]. Since the steady state is rapidly established,
the transfer-matrix approach is well applicable in this
no-saturation regime. Absence of saturation is directly demonstrated
by the inset in Fig. \ref{fig4}, where the population difference
does not change during establishment of the steady state. The
initial transient regime is unavoidable in realistic systems. It can
be studied with the FDTD method, and $\mathcal{PT}$ symmetry is
unreachable in this mode due to impossibility to change the sign of
relaxation rates $\gamma_1$ and $\gamma_2$ (see Ref. \cite{Supp}).

\section{\label{trans}Phase transition dynamics}

\begin{figure}[t!]
\includegraphics[scale=1., clip=]{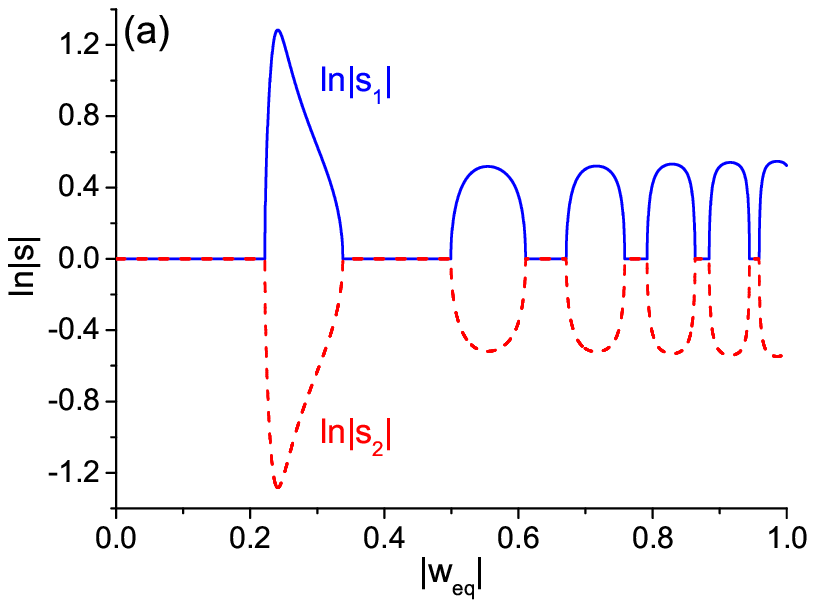}
\includegraphics[scale=1., clip=]{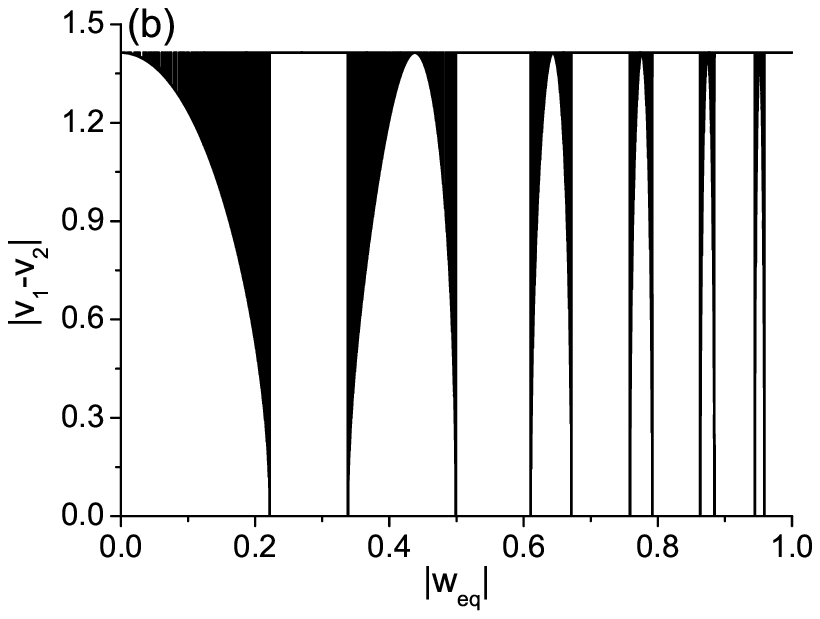}
\centering \caption{\label{fig5} (a) Logarithm of eigenvalues $s_1$
and $s_2$ and (b) difference of eigenvectors $|{\bf v}_1-{\bf v}_2|$
of the scattering matrix as a function of the pumping parameter
$|w_{eq}|$.}
\end{figure}

$\mathcal{PT}$-symmetry breaking can be considered as a peculiar
phase transition. It can be realized either by changing wavelength
$\lambda$ of light for a given value of pumping parameter
$|w_{eq}|$, or, conversely, by changing the pumping parameter at a
fixed wavelength. The latter variant is analyzed in this paper. The
former one deserves a separate study, since it implies the necessity
to consider the effects of frequency detuning. The criterion of the
$\mathcal{PT}$-symmetry breaking is usually formulated in terms of
the eigenvalues $s_1$ and $s_2$ of the scattering matrix
\cite{Ge2012}. The scattering matrix connects the left and right
input fields with left and right output fields. A
$\mathcal{PT}$-symmetric system has unimodular eigenvalues
$|s_1|=|s_2|=1$ of the scattering matrix. When $\mathcal{PT}$
symmetry is violated, the modules of the eigenvalues are inverse as
$|s_1|>1$ and $|s_2| = 1/|s_1| < 1$. At the points of phase
transition called exceptional points, the eigenvectors of the
scattering matrix coincide. Using the definition of the scattering
matrix \cite{Ge2012}
\begin{eqnarray}
S=\left( \begin{array}{cc}{r_{LG} \qquad t_{GL}} \\ {t_{LG} \qquad
r_{GL}} \end{array} \right), \label{scat}
\end{eqnarray}
we calculate both eigenvalues and eigenvectors for pumping parameter
$|w_{eq}|$ swept through the whole interval from $0$ to $1$. Here
$t_{LG}$, $t_{GL}$, $r_{LG}$, and $r_{GL}$ are the transmission and
reflection coefficients, which can be expressed through the
respective elements of transfer matrix $M$: $t_{LG}=1/M_{11}$,
$t_{GL}=\det[M]/M_{11}=t_{LG}$ (since $\det[M]=1$),
$r_{LG}=M_{21}/M_{11}$, and $r_{GL}=-M_{12}/M_{11}$. Stationary
transmission and reflection are calculated as
$T=|t_{LG}|^2=|t_{GL}|^2$, $R_{LG}=|r_{LG}|^2$, and
$R_{GL}=|r_{GL}|^2$. Results of transfer-matrix calculations for the
multilayer structure with the permittivities (\ref{epsPT}) are shown
in Fig. \ref{fig5}. At the first exceptional point $|w_{eq}| \approx
0.222$, the phase transition occurs and eigenvalues cease to be
unimodular [Fig. \ref{fig5}(a)], whereas the difference between
eigenvectors vanishes [Fig. \ref{fig5}(b)]. Unimodularity is
violated and, therefore, $\mathcal{PT}$ symmetry is broken up to the
second exceptional point. The latter returns the system into the
$\mathcal{PT}$-symmetric state. All in all, there is a number of
ranges of $|w_{eq}|$ with broken symmetry and a corresponding number
of exceptional points.

\begin{figure}[t!]
\includegraphics[scale=0.95, clip=]{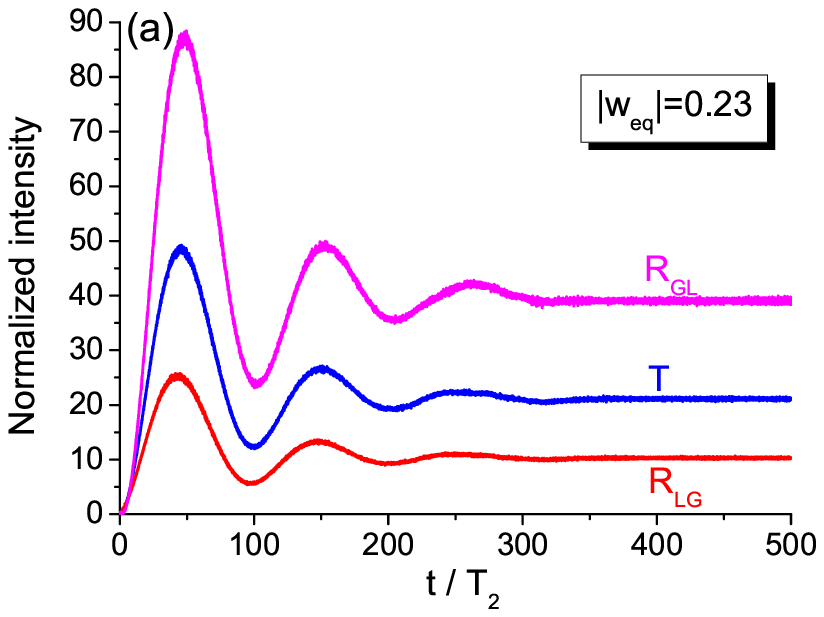}
\includegraphics[scale=0.95, clip=]{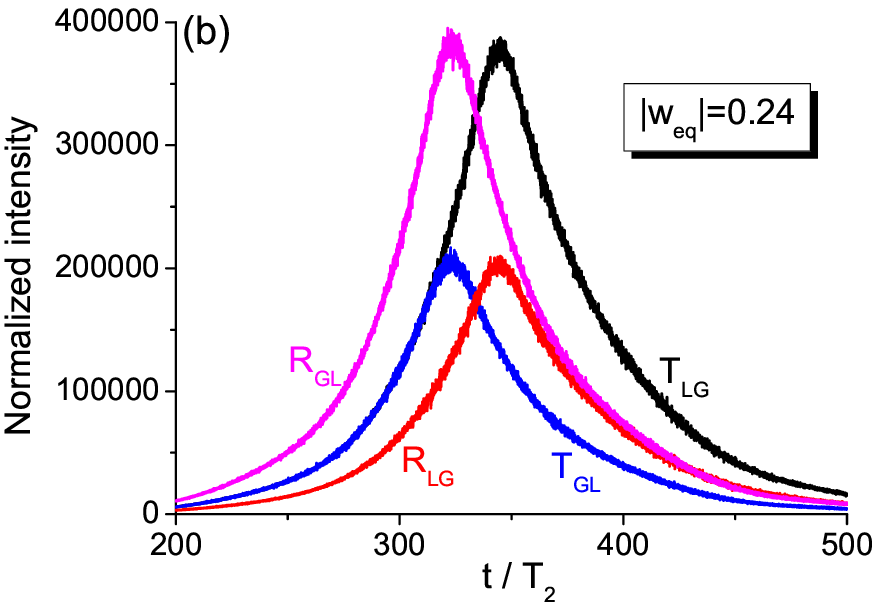}
\centering \caption{\label{fig6} Temporal dynamics of the reflected
(R) and transmitted (T) intensity for the pumping parameter (a)
$|w_{eq}|=0.23$ and (b) $|w_{eq}|=0.24$.}
\end{figure}

Transfer-matrix approach used so far cannot describe dynamics of the
wave propagation in resonant media by its definition. Now we will
employ the rigorous Maxwell-Bloch equations for studying light
dynamics near an exceptional point. There is a dramatic discrepancy
between the two calculation techniques, when the pumping parameter
approaches the exceptional point. Relative difference in
transmission calculated with help of the Maxwell-Bloch and
transfer-matrix approaches, $|T_{MB}-T_{TM}|/T_{MB} \approx 0.13$,
results in the satisfactory agreement for $|w_{eq}|=0.22$, but its
value $0.48$ at $|w_{eq}|=0.23$ is unacceptably large. Such a large
discrepancy stems from the qualitatively different behaviors of the
system at $|w_{eq}|=0.23$: the system is above the exceptional point
according to the transfer-matrix method, whereas it is still in the
$\mathcal{PT}$-symmetric state according to the FDTD simulations. In
fact, temporal dynamics with an established stationary state in Fig.
\ref{fig6}(a) is distinctive for the $\mathcal{PT}$ symmetry (cf.
Fig. \ref{fig4}). Formation of the steady state after a rather short
time corroborates existence of the balance between gain and loss at
$|w_{eq}|=0.23$. The general tendency is that the closer to the
exceptional point, the longer the transient period is. At the
exceptional point [Fig. \ref{fig6}(b)], the field rapidly grows
changing population difference $w$, this grow being limited by
saturation.

\begin{figure*}[t!]
\includegraphics[scale=1.5, clip=]{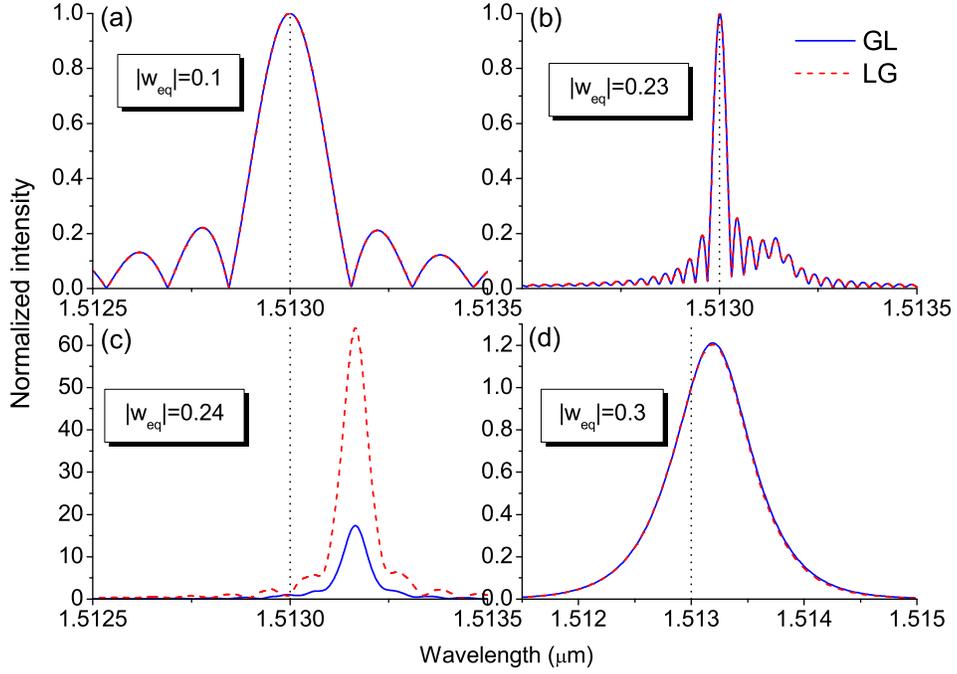}
\centering\caption{\label{fig7} Spectra of transmitted radiation for
different levels of pumping: (a) $|w_{eq}|=0.1$, (b)
$|w_{eq}|=0.23$, (c) $|w_{eq}|=0.24$, and (d) $|w_{eq}|=0.3$. The
spectra are normalized on the intensity at the resonant wavelength
$1.513$ $\mu$m.}
\end{figure*}

Breaking of $\mathcal{PT}$ symmetry is expected to result in the
strong (exponential) amplification of propagating waves due to the
fact that gain can not be compensated with losses in this case. In
Fig. \ref{fig6}(b), instability exhibiting very strong light
amplification is observed at $|w_{eq}|=0.24$. The energy pumped in
the system is promptly released as a high-intensity pulse. In
concordance with Ref. \cite{Novitsky2017}, this regime can be called
a lasing-like (or quasilasing) mode. It can be treated as a
dynamical feature of the phase state of broken $\mathcal{PT}$
symmetry. In contrast to the true lasing the pulse is generated not
by small field fluctuations, but rather in response to the incident
wave (though the intensity of input wave can be taken much lower, as
evidenced by Supplemental Material Fig. S4 \cite{Supp}). At the
exceptional point, we have a very long transient period and the
stationary levels of reflection and transmission are again expected
to occur in the long-time limit \cite{Novitsky2017}. These levels
strongly differ from those calculated with the transfer-matrix
method due to saturation development discussed further. In other
words, the resulting population difference will be no more
determined by pumping parameter $w_{eq}$ as assumed in Eq.
(\ref{epsPT}). The pulse gets shorter and more powerful with
increasing $|w_{eq}|$ as evidenced by comparison of Fig.
\ref{fig6}(b) and Supplemental Material Fig. S1 \cite{Supp}. The
spectra of transmitted radiation shown in Fig. \ref{fig7} indicate
the shift of maximal amplification from the resonant wavelength
$\lambda=1.513$ $\mu$m (below the exceptional point) to longer
wavelengths (above the exceptional point). This redshift in the
lasing-like regime can be explained by two factors: (i) higher light
absorption on the resonant wavelength in the loss layers, so that
amplification at the neighboring wavelengths becomes prevalent, (ii)
intensity modulation due to incomplete stationary-state
establishment (see Supplemental Material Fig. S5 \cite{Supp}). We
should stress that although the matrix method with permittivities
(\ref{epsPT}) is able to approximately determine an exceptional
point, it fails in adequate description of the phase transition and
in describing temporal dynamics of light-structure interaction. In
other words, the full system of Maxwell-Bloch equations should be
exploited in the vicinity of the points of $\mathcal{PT}$-symmetry
breaking.

\begin{figure*}[t!]
\includegraphics[scale=1.5, clip=]{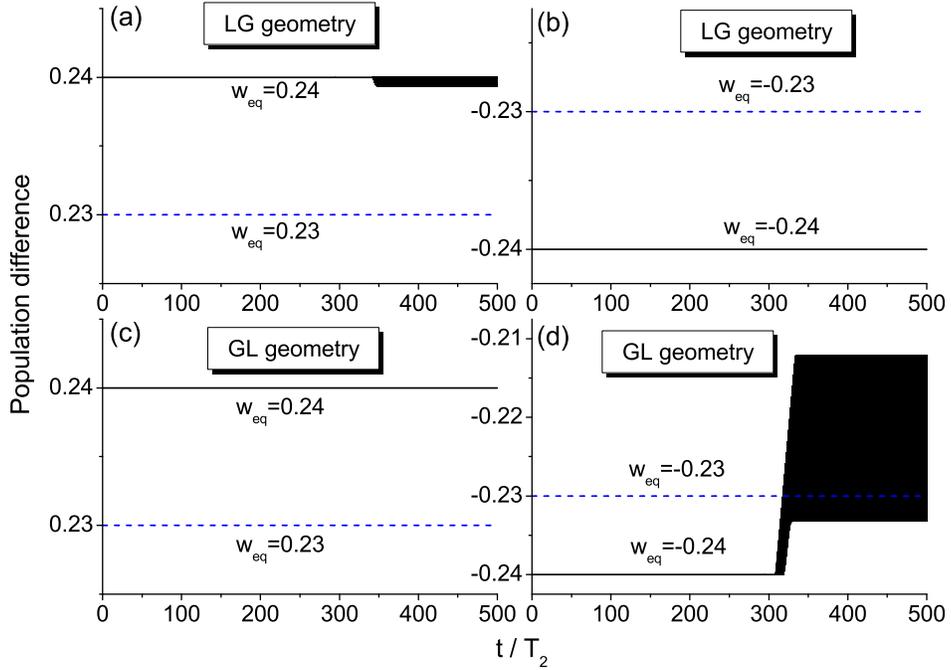}
\centering \caption{\label{fig8} Temporal dynamics of the population
difference in the first unit cell of LG and GL structures at the
pumping parameters $|w_{eq}|=0.23$ and $0.24$: (a), (c) $w(t)$ in
the loss layer and (b), (d) $w(t)$ in the gain layer. }
\end{figure*}

Strong amplification of a signal in the non-$\mathcal{PT}$-symmetric
phase [Fig. \ref{fig6}(b)] plays important role in loss and gain
saturation. Indeed, normalized light amplitude $\Omega$ inside the
system is not much less than $\Omega_{sat}$ anymore. Population
difference preserves its initial value $w(t)=w_{eq}$ in the
$\mathcal{PT}$-symmetric state below the exceptional point (see the
dashed lines in Fig. \ref{fig8}, for $|w_{eq}|=0.23$). However,
fluctuations of the population difference due to saturation occur
above the exceptional point, at $|w_{eq}|=0.24$, in both LG (in the
loss layer) and GL (in the gain layer) configurations. Saturation
imposes constraint on further increase of light intensity as
evidenced by the coincidence of the intensity peak in Fig.
\ref{fig6}(b) and saturation development in time in Fig. \ref{fig8}.
Fluctuations indicate that the $\mathcal{PT}$ symmetry is broken
through violation of the necessary condition $\varepsilon(z) =
\varepsilon^\ast(-z)$ and loss and gain are not balanced anymore.
Saturation also leads to \textit{the irreversible phase transition}
in the system: the return of the system to the
$\mathcal{PT}$-symmetric state predicted by the stationary theory at
larger pumping parameters (see Fig. \ref{fig5}) is impossible, since
Eq. (\ref{epsPT}) is not valid anymore. Direct FDTD calculations for
$|w_{eq}| > 0.24$ (see Supplemental Material Fig. S1 \cite{Supp}) do
support this conclusion resulting in the lasing-like dynamics
similar to that shown in Fig. \ref{fig6}(b).

Broken $\mathcal{PT}$ symmetry drastically affects dynamics of the
transmitted and reflected waves. Owing to the nonlinear process of
saturation, the transmission becomes asymmetric, $T_{LG} \neq
T_{GL}$, i.e., the multilayer structure is \textit{nonreciprocal}.
Usually the saturation-induced nonreciprocity is introduced through
the nonlinear permittivity Eq. (\ref{epsTLM})
\cite{Liu2014,Barton2017}, but solution of dynamic Eqs.
(\ref{dPdtau})-(\ref{Maxdl}) is more accurate and informative. In
the saturation regime the system is non-Hermitian, but it can be
linearized to a $\mathcal{PT}$-symmetric multilayer
\cite{Barton2017}.

Intensities of the pulses escaping the system do not depend on the
direction of incident light: almost the same pulses are emitted from
the gain and loss ends of the multilayer after reversing the input
light direction ($T_{LG} = R_{GL}$ and $T_{GL} = R_{LG}$) as shown
in Fig. \ref{fig6}(b). In other words, direction of the output
pulses is \textit{locked by $\mathcal{PT}$-symmetry breaking}. This
locking can be presumably caught only within the dynamical
calculations, because it has not been reported earlier.
Nonreciprocal transmission is accompanied by the propagation
direction locking at higher pumping as well, what is demonstrated in
Supplemental Material Fig. S1 \cite{Supp} for $|w_{eq}|=0.3$ and
$0.4$. Locking of the light propagation directions can be viewed as
a possible basis for peculiar all-optical diodes and transistors.

\begin{figure}[t!]
\includegraphics[scale=0.9, clip=]{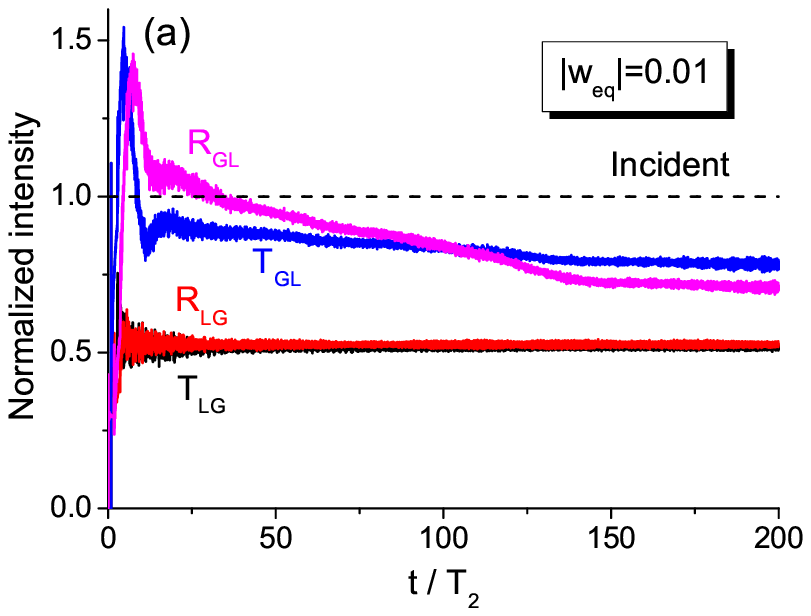}
\includegraphics[scale=0.9, clip=]{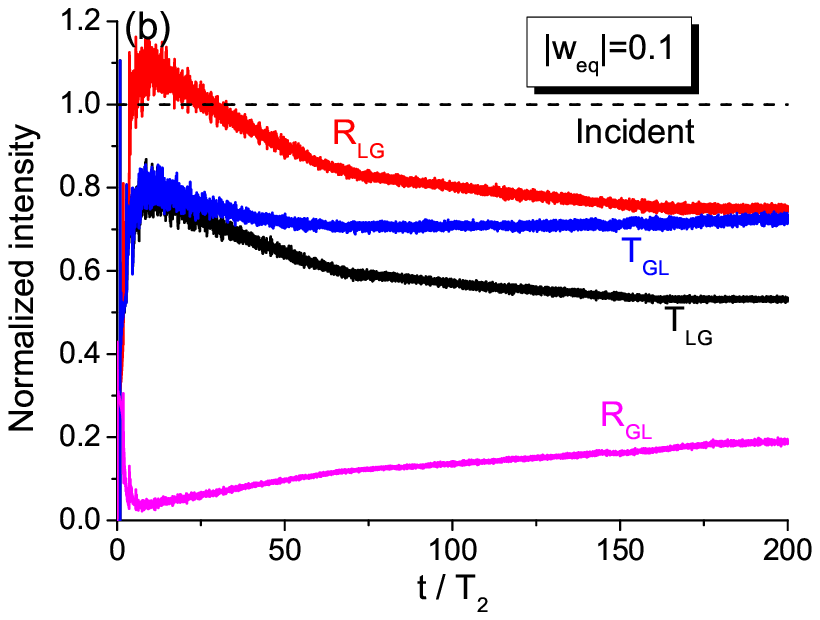}
\centering \caption{\label{fig9} Temporal dynamics of the reflected
(R) and transmitted (T) intensity for the pumping parameter (a)
$|w_{eq}|=0.01$ and (b) $|w_{eq}|=0.1$ in the case of the wave with
relatively large initial amplitude $\Omega_0=10^{-2} \gamma_2$.}
\end{figure}

It should be emphasized that the saturation is not the reason for
this locking. In order to demonstrate this, we consider wave
propagation in saturation regime (for initial amplitude
$\Omega_0=10^{-2} \gamma_2$) which breaks $\mathcal{PT}$ symmetry at
every value of the pumping parameter. Nonreciprocity of transmission
due to saturation is clearly seen in Fig. \ref{fig9}, but the
direction locking of output pulses is missing. Therefore, the
$\mathcal{PT}$-symmetry breaking is necessary for this locking to
occur.

The behavior similar to that described above generally occurs near
the exceptional points as evidenced by Figs. S2, S3, and S4 in
Supplemental Material \cite{Supp}. In particular, the number of
layers controls position of the exceptional point: when the
structure length is decreased, higher pumping is needed for
$\mathcal{PT}$-symmetry breaking (Fig. S2 of Supplemental Material
\cite{Supp}) and vice versa. Although system's response is generally
very complex due to interplay of the loss/gain and multilayer
resonances \cite{Witonski2017}, the results similar to discussed
above are expected for other parameters of the structure and proper
operating frequency. Such scalability is encouraging for designing
realistic realizations of $\mathcal{PT}$-symmetric multilayers. We
would like to emphasize that another model of
$\mathcal{PT}$-symmetric multilayer with unexcited absorbing layers
mentioned in Section \ref{eqpars} provides the similar results (see
Supplemental Material Fig. S6 \cite{Supp}).

\section{\label{concl}Conclusion}

We have analyzed temporal dynamics of light in
$\mathcal{PT}$-symmetric periodic multilayers, the gain and loss
slabs being modeled as a resonant medium. Light-matter interactions
in the resonant media are described by the Maxwell-Bloch equations,
which are simulated numerically to provide deeper insight into
transition dynamics between the $\mathcal{PT}$-symmetric and
$\mathcal{PT}$-broken phases. In particular, predictions of the
stationary transfer-matrix method are shown to be inadequate in the
vicinity of the exceptional points. We feature the so-called
lasing-like regime in the $\mathcal{PT}$-symmetry broken state
characterized by emission of powerful pulses of radiation and
development of saturation. The latter is the reason for phase
transition irreversibility -- that is, the system cannot return to
the $\mathcal{PT}$-symmetric state for the pumping parameters above
the exceptional point. In the $\mathcal{PT}$-broken phase, the
direction of pulses escaping the system is found to be locked by the
$\mathcal{PT}$-symmetry breaking, meaning that the intensities of
two output waves are independent of the direction of the incident
radiation. The approach based on the Maxwell-Bloch equations seems
to be rather general and applicable to structures with other
geometries, e.g., coupled ring resonators \cite{Feng2014}. We
envisage its application to investigation of other effects near the
exceptional points, such as coherent perfect absorption
(anti-lasing) \cite{Wong2016,Chong2010,Longhi2010}. Intricate
interplay between loss and gain in $\mathcal{PT}$-symmetric systems
opens up new opportunities for constructing photonic devices for
optical communications, computing, and sensing. The approach
proposed here is expected to be useful to realize some of these
diverse functionalities.

\acknowledgements{The work was supported by the Belarusian
Republican Foundation for Fundamental Research (Projects No.
F16K-016 and F18R-021), the Russian Foundation for Basic Research
(Projects No. 18-02-00414, 18-52-00005 and 18-32-00160), Ministry of
Education and Science of the Russian Federation (GOSZADANIE, Grant
No. 3.4982.2017/6.7), Government of Russian Federation (Grant
08-08), and the Israeli Ministry of Trade and Labor Kamin Program
(Grant No. 62045). Numerical simulations of light interaction with
resonant media were supported by the Russian Science Foundation
(Project No. 17-72-10098). The calculations of field distributions
were supported by the Russian Science Foundation (Project No.
16-12-10287). Partial financial support of Villum Fonden (DarkSILD
project) is acknowledged.}

\end{document}


\title{Supplemental Material for \\ $\mathcal{PT}$-symmetry breaking in multilayers with resonant loss and gain locks light propagation direction}

\author{Denis~V.~Novitsky$^{1,2,3}$}
\author{Alina~Karabchevsky$^{4,5,6}$}
\author{Andrei~V.~Lavrinenko$^7$}
\author{Alexander~S.~Shalin$^2$}
\author{Andrey~V.~Novitsky$^{7,8}$}

\affiliation{$^1$B.~I.~Stepanov Institute of Physics, National
Academy of Sciences of Belarus, Nezavisimosti Avenue 68, 220072
Minsk, Belarus \\ $^2$ITMO University, Kronversky Prospekt 49,
197101 St. Petersburg, Russia \\ $^3$Saint Petersburg
Electrotechnical University ``LETI'', 5 Prof. Popova Str., 197376
St. Petersburg, Russia \\ $^4$ Electrooptical Engineering Unit,
Ben-Gurion University of the Negev, David Ben-Gurion Blvd, P.O.B.
653, 8410501 Beer-Sheva, Israel \\ $^5$ Ilse Katz Institute for
Nanoscale Science $\&$ Technology, Ben-Gurion University of the
Negev, Beer-Sheva, 8410501, Israel \\ $^6$ Center for Quantum
Information Science and Technology, Ben-Gurion University of the
Negev, Beer-Sheva, 8410501, Israel \\ $^7$ DTU Fotonik, Technical
University of Denmark, {\O}rsteds Plads 343, DK-2800 Kongens Lyngby,
Denmark \\ $^8$ Department of Theoretical Physics and Astrophysics,
Belarusian State University, Nezavisimosti Avenue 4, 220030 Minsk,
Belarus}

\begin{abstract}
\end{abstract}

\maketitle

\section{Derivation of the conditions for $\mathcal{PT}$ symmetry}

Let us derive the conditions for $\mathcal{PT}$ symmetry directly
from the Maxwell-Bloch equations. We start with the
quantum-mechanics-like equation
\begin{equation}
\hat H({\bf r}, t) \psi({\bf r}, t) = \lambda \psi({\bf r}, t),
\label{H}
\end{equation}
where $\hat H$ is a differential operator. It is transformed to
\begin{equation}
\hat H^\ast(-{\bf r}, -t) \psi^\ast(-{\bf r}, -t) = \lambda^\ast
\psi^\ast(-{\bf r}, -t), \label{HPT}
\end{equation}
under the action of parity $\hat P$ and time $\hat T$ operators. In
other words, the $\hat P \hat T$ operator changes $z \rightarrow
-z$, $t \rightarrow -t$, and $i \rightarrow -i$. The system is
$\mathcal{PT}$-symmetric, if $\hat H({\bf r}, t) = \hat H^\ast(-{\bf
r}, -t)$ and $\psi({\bf r}, t) = \psi^\ast(-{\bf r}, -t)$. Then Eqs.
(\ref{H}) and (\ref{HPT}) coincide and $\hat H$ has a real spectrum
$\lambda = \lambda^\ast$.

Now we turn to the Maxwell-Bloch equations and denote the action of
$\hat P \hat T$-operator with the tilde, e.g., $\tilde \rho = \hat P
\hat T \rho(z,t) = \rho^\ast(-z,-t)$. Acting with the $\mathcal{PT}$
operator on Eqs. (1)-(3) of the main text, one gets
\begin{eqnarray}
\frac{d\tilde\rho}{d\tau}&=& i \tilde l \tilde\Omega \tilde w + i \tilde\rho \tilde\delta + \gamma_2 \tilde\rho, \label{dPdtauPT} \\
\frac{d\tilde w}{d\tau}&=&2 i (\tilde l^* \tilde\Omega^* \tilde\rho
- \tilde\rho^* \tilde l \tilde\Omega) + \gamma_1 (\tilde w - \tilde
w_{eq}),
\label{dNdtauPT} \\
\frac{\partial^2 \tilde\Omega}{\partial \xi^2}&-& \tilde n_d^2
\frac{\partial^2 \tilde\Omega}{\partial \tau^2}+2 i \frac{\partial
\tilde\Omega}{\partial \xi}+2 i \tilde n_d^2 \frac{\partial
\tilde\Omega}{\partial
\tau} + (\tilde n_d^2-1) \Omega \nonumber \\
&&=3 \tilde\alpha \tilde l \left(\frac{\partial^2
\tilde\rho}{\partial \tau^2}-2 i \frac{\partial \tilde\rho}{\partial
\tau} - \tilde\rho\right). \label{MaxdlPT}
\end{eqnarray}

Eq. (\ref{MaxdlPT}) coincides with Eq. (3) from the main text, if
$n_d = \tilde n_d$, $\alpha = \tilde\alpha$, $\Omega = \tilde
\Omega$, and $\rho = \tilde \rho$. In other words, $n_d(\xi) =
n_d^\ast(-\xi)$, $\alpha(\xi) = \alpha(-\xi)$, $\Omega(\xi,\tau) =
\Omega^\ast(-\xi,-\tau)$, and $\rho(\xi,\tau) =
\rho^\ast(-\xi,-\tau)$. Condition $n_d(\xi) = n_d^\ast(-\xi)$ leads
to $l(\xi) = l^\ast(-\xi)$. The relaxation rates cannot change the
sign, therefore, Eqs. (\ref{dPdtauPT}) and (\ref{dNdtauPT}) cannot
be turned into Eqs. (1) and (2) of the main text. Thus, the
$\mathcal{PT}$ symmetry is not applicable to the Maxwell-Bloch
equations in general, but only to the particular cases of negligible
relaxation rates (in comparison with the frequency of light) and/or
stationary regime.

In the stationary regime, time-independent $\Omega$, $w$ and $\rho$
can be joined into a vector $\psi(\xi) = (\Omega, \rho, W)^T$, where
$W = i w$, and Eqs. (1) and (2) of the main text read
\begin{eqnarray}
&& l \Omega W + i \rho \delta - \gamma_2 \rho = 0, \nonumber \\
&& l^* \Omega^* \rho - \rho^* l \Omega + \gamma_1 (W - i w_{eq}) =
0.
\end{eqnarray}
Operator $\hat P \hat T$ transforms these equations to
\begin{eqnarray}
&& \tilde l \tilde\Omega \tilde W - i \tilde\rho \tilde\delta - \gamma_2 \tilde\rho = 0, \nonumber \\
&& \tilde l^* \tilde\Omega^* \tilde\rho - \tilde\rho^* \tilde l
\tilde\Omega + \gamma_1 (\tilde W + i \tilde w_{eq}) = 0.
\end{eqnarray}

PT-symmetry requires $\psi = \tilde \psi$, $w_{eq} = -\tilde
w_{eq}$, and $\delta = -\tilde\delta$, or $\psi(\xi) =
\psi^\ast(-\xi)$, $w_{eq}(\xi) = -w_{eq}(-\xi)$, and $\delta(\xi) =
-\delta(-\xi)$.

Thus, conditions for existence of $\mathcal{PT}$ symmetry in
resonant media are (i) the stationary regime, (ii) $w_{eq}(\xi) =
-w_{eq}(-\xi)$, and (iii) $\delta(\xi) = -\delta(-\xi)$. The same
conditions follow from Eqs. (4) and (5).

Note also that in the case of the low-intensity incident pulse (as
compared to the saturation intensity, $\Omega << \Omega_{sat}$), the
population difference $w$ remains unchanged and equal to $w_{eq}$,
while the polarization $\rho \sim \Omega$ remains small and rapidly
takes the stationary value. Under such conditions, the system
remains \textit{almost} $\mathcal{PT}$-symmetric during all the
process of the stationary-state establishment.

\section{Calculations for the case discussed in the main text}

\begin{figure*}[b]
\centering \includegraphics[scale=1.0,clip=]{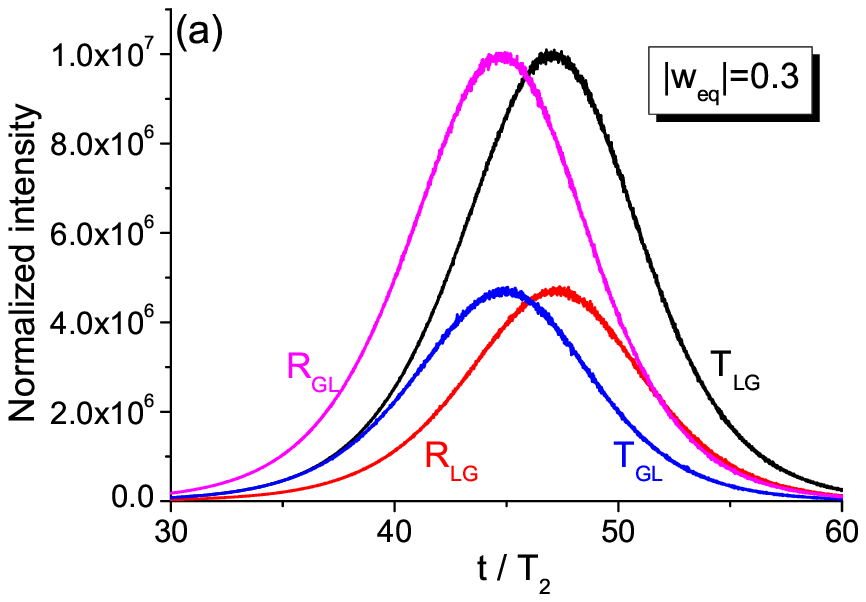}
\includegraphics[scale=1.0,clip=]{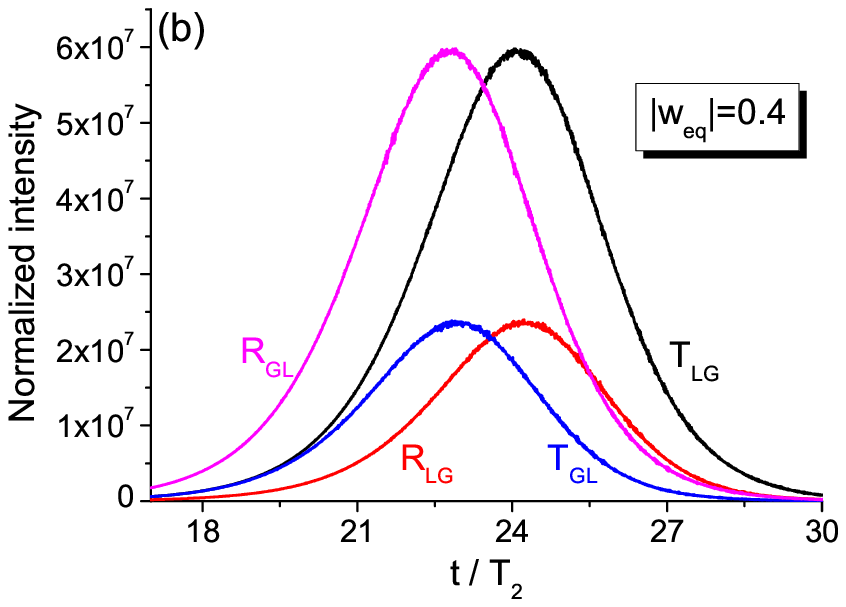}
\caption{Temporal dynamics of the reflected (R) and transmitted (T)
intensity for the pumping parameter (a) $|w_{eq}|=0.3$ and (b)
$|w_{eq}|=0.4$.} \label{figS1}
\end{figure*}

In this section, we give results of additional calculations which
support Section 4 of the main text. Figure \ref{figS1} shows the
profiles of reflected and transmitted intensity for the pumping
parameters $|w_{eq}|=0.3$ and $0.4$ at the same parameters as in
Fig. 6. These profiles possess the same pattern as in Fig. 6(b),
i.e., we see direction locking and nonreciprocity characteristic for
the lasing-like regime, but the pulses are much shorter and more
smooth. Note that scattering-matrix calculations predict
$\mathcal{PT}$-symmetric state at $|w_{eq}|=0.4$ (see Fig. 5(a)).


\begin{figure*}[t]
\centering \includegraphics[scale=1.0,clip=]{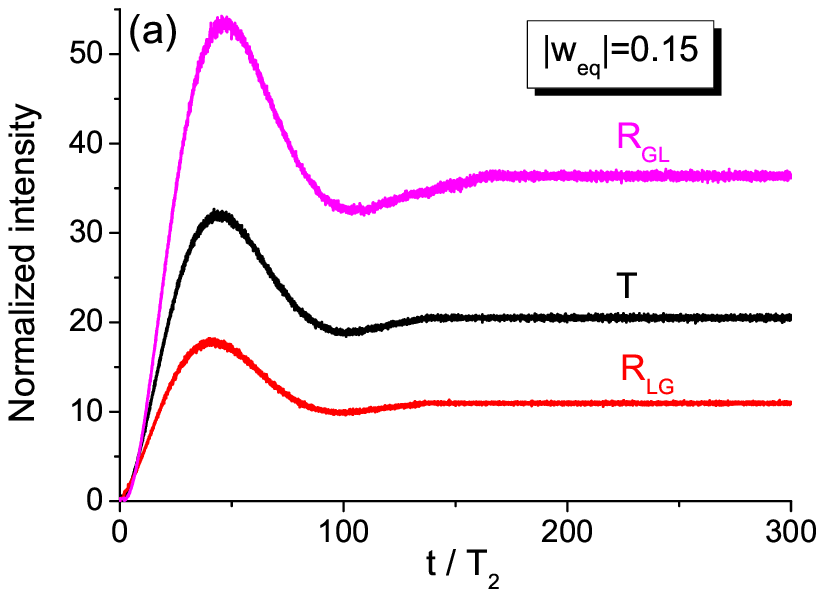}
\includegraphics[scale=1.0,clip=]{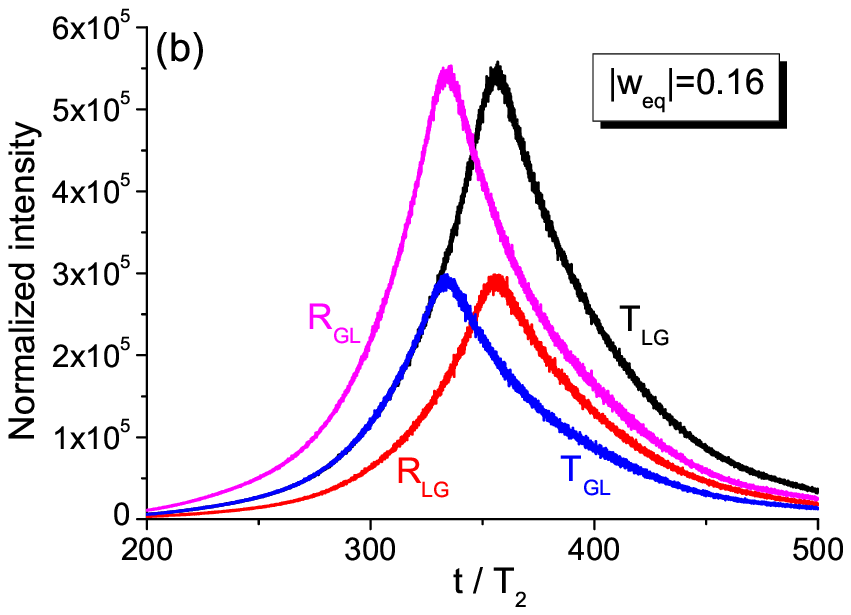}
\includegraphics[scale=1.0,clip=]{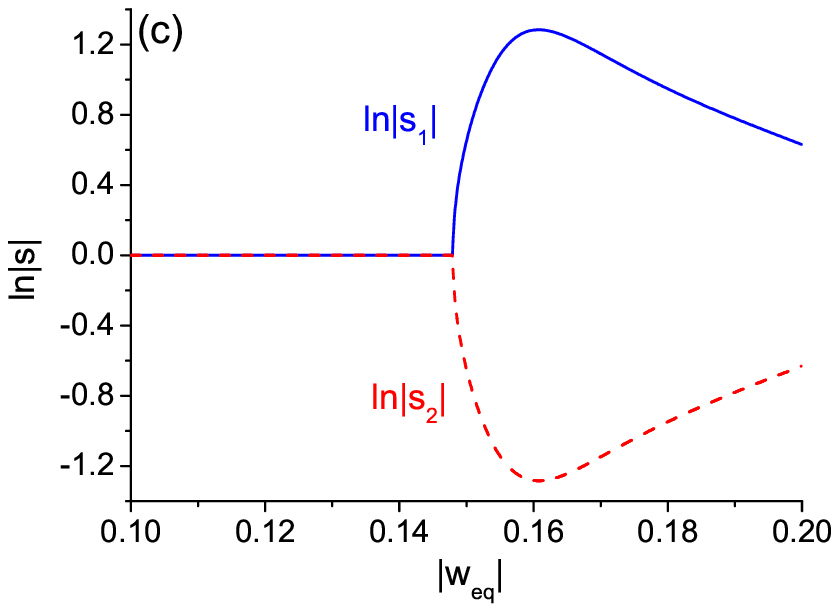}
\caption{The case of increased particle density ($\omega_L=1.5
\times 10^{11}$ s$^{-1}$): Temporal dynamics of the reflected (R)
and transmitted (T) intensity for the pumping parameter (a)
$|w_{eq}|=0.15$ and (b) $|w_{eq}|=0.16$. (c) Corresponding
dependence of logarithm of the scattering-matrix eigenvalues $s_1$
and $s_2$ on the pumping parameter. The eigenvalues are calculated
with the transfer-matrix method in stationary approximation.}
\label{figS2}
\end{figure*}


\begin{figure*}[t]
\centering \includegraphics[scale=1.0,clip=]{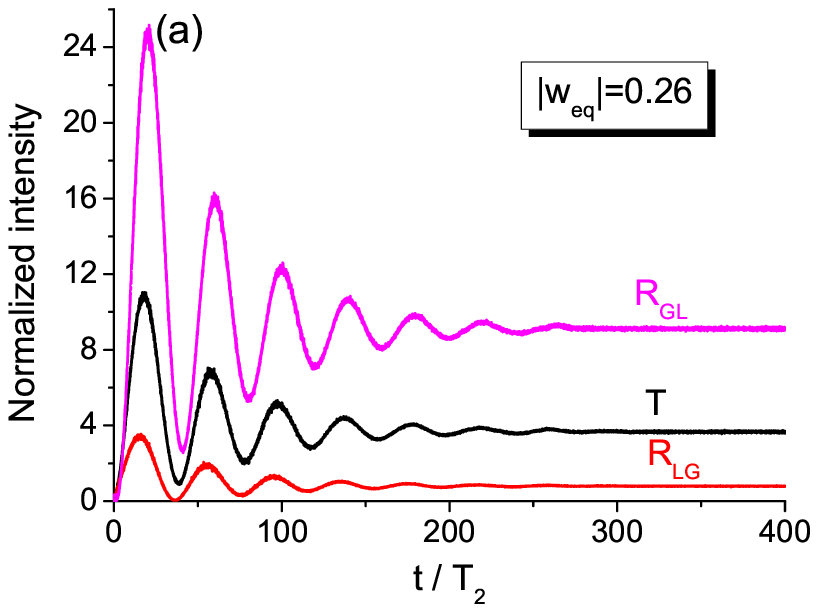}
\includegraphics[scale=1.0,clip=]{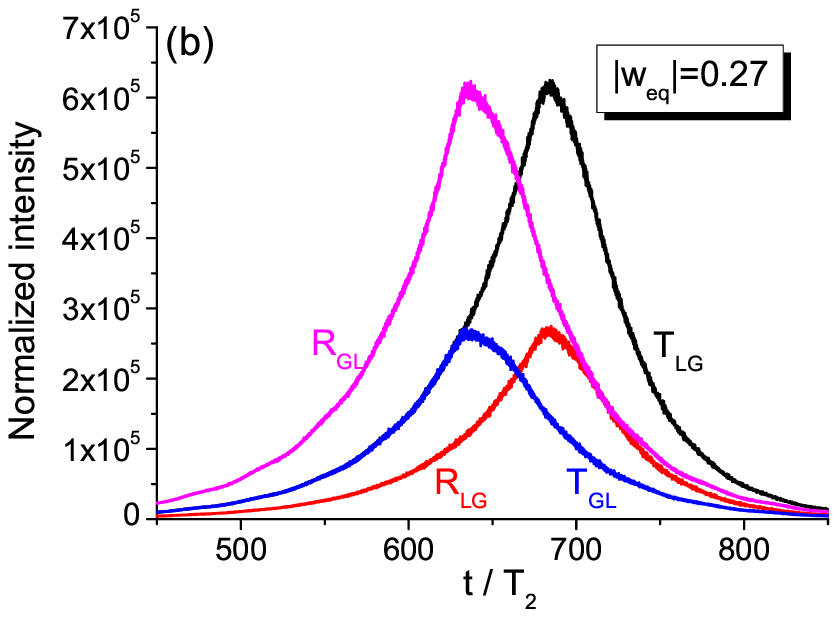}
\includegraphics[scale=1.0,clip=]{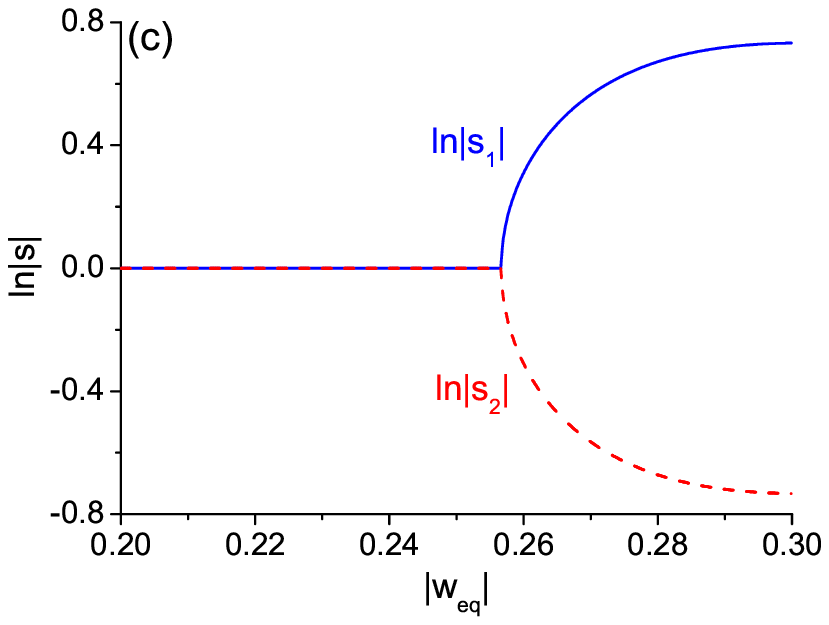}
\caption{The case of decreased number of layers ($N=15$): Temporal
dynamics of the reflected (R) and transmitted (T) intensity for the
pumping parameter (a) $|w_{eq}|=0.26$ and (b) $|w_{eq}|=0.27$. (c)
Corresponding dependence of logarithm of the scattering-matrix
eigenvalues $s_1$ and $s_2$ on the pumping parameter. The
eigenvalues are calculated with the transfer-matrix method in
stationary approximation.} \label{figS3}
\end{figure*}


\begin{figure*}[t]
\centering \includegraphics[scale=1.0,clip=]{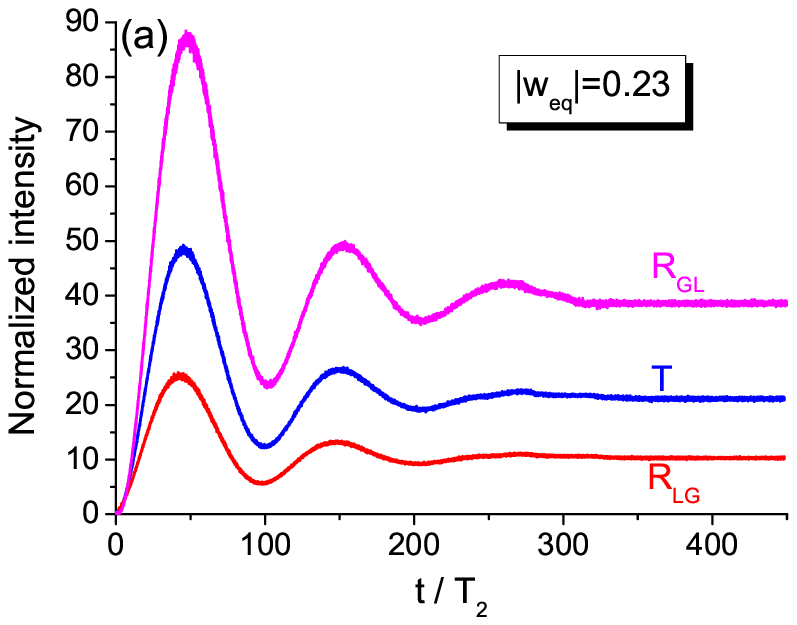}
\includegraphics[scale=1.0,clip=]{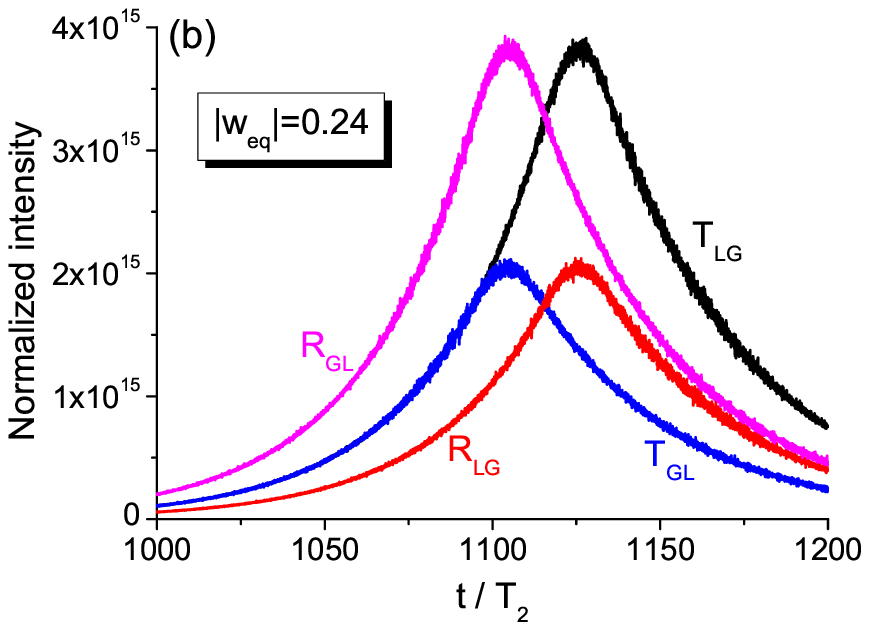}
\caption{The case of lower input intensity $\Omega_0 = 10^{-10}
\gamma_2$: Temporal dynamics of the reflected (R) and transmitted
(T) intensity for the pumping parameter (a) $|w_{eq}|=0.23$ and (b)
$|w_{eq}|=0.24$.} \label{figS4}
\end{figure*}


\begin{figure*}[t]
\centering \includegraphics[scale=2,clip=]{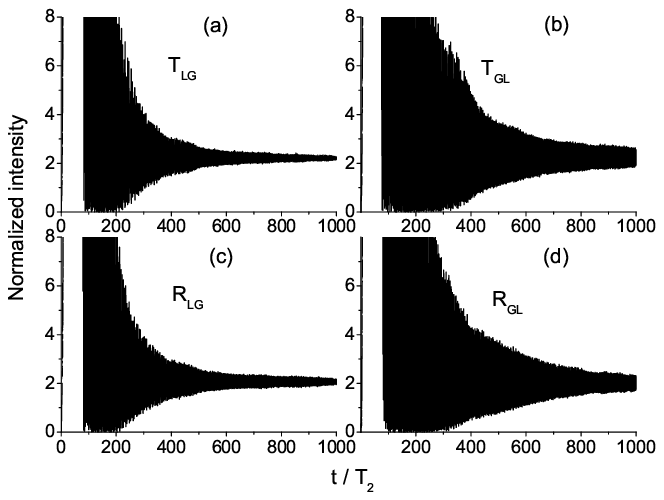}
\caption{Stationary state establishment for $|w_{eq}|=0.3$ [see Fig.
\ref{figS1}(a)].} \label{figS5}
\end{figure*}


In order to prove that the lasing-like regime is connected with
$\mathcal{PT}$-symmetry breaking, we calculate wave propagation
through the structures with the changed parameters. This change
results in the shift of the exceptional point. Figure \ref{figS2}
shows the case of $1.5$-times increased density of active particles,
so that the Lorentz frequency is $\omega_L=1.5 \times 10^{11}$
s$^{-1}$. In this case, $\mathcal{PT}$-symmetry breaking occurs at
lower pumping parameters in comparison with the case discussed in
the main text. On the contrary, decreasing the number of layers in
the structure leads to the shift of the exceptional point to larger
$|w_{eq}|$ as seen in Fig. \ref{figS3} for $N=15$. In both cases,
however, breaking of $\mathcal{PT}$ symmetry is accompanied by the
transition to the lasing-like regime with direction locking clearly
seen.

We have also studied the case of much lower incident wave amplitude
$\Omega_0 = 10^{-10} \gamma_2$ than that considered in the main text
($\Omega_0 = 10^{-5} \gamma_2$). The results for this case shown in
Fig. \ref{figS4} indicate that the position of exceptional point
does not depend on initial light intensity and, hence, is the
inherent characteristic of the system. \textit{Relative} (or
normalized by the incident intensity) transmission and reflection in
Fig. \ref{figS4}(a) and Fig. 6(a) are identical in the
$\mathcal{PT}$-symmetric state. In the state of broken
$\mathcal{PT}$ symmetry, the intensities of transmitted and
reflected pulses in Fig. \ref{figS4}(b) and Fig. 6(b) have the same
\textit{absolute} (non-normalized) values reached over different
periods: the weaker input wave, the longer period.

Finally, Fig. \ref{figS5} shows the process of steady-state
establishment above the exceptional point. One can see that this
process is much longer than in the $\mathcal{PT}$-symmetric state
and is accompanied by violent fluctuations of intensity, which may
be one of the reasons for spectral maximum shift reported in Fig. 7
of the main text.

\section{Calculations for the case of completely unexcited absorbing layers}

This section contains additional information on another possibility
to achieve $\mathcal{PT}$ symmetry mentioned in Section 2 of the
main text. We take completely unexcited absorbing layers
($w_{eq,L}=1$) and pump the amplifying layers to a certain level
$w_{eq,G}<0$, which depends on the relationship between densities
$C$ of active particles in loss and gain layers (or, equivalently,
on the relationship between the light-matter coupling coefficients,
since $\alpha \sim C$). Indeed, for the loss layers we have
\begin{eqnarray}
\varepsilon_{eff+} \approx n_d^2 + 3 i l^2 \omega_{L,L} T_2,
\label{epsloss}
\end{eqnarray}
whereas for the gain layers we can write
\begin{eqnarray}
\varepsilon_{eff-} \approx n_d^2 - 3 i l^2 \omega_{L,G} T_2
|w_{eq,G}|. \label{epsgain}
\end{eqnarray}
The necessary condition for $\mathcal{PT}$ symmetry requires the
equality of imaginary parts Eqs. (\ref{epsloss}) and
(\ref{epsgain}), so that $|w_{eq,G}| = \omega_{L,L}/\omega_{L,G} =
\alpha_{L}/\alpha_{G}$, where $\alpha_{L}$ and $\alpha_{G}$ are
light-matter coupling coefficients for loss and gain layers,
respectively.

\begin{figure*}[t]
\centering \includegraphics[scale=1.0,clip=]{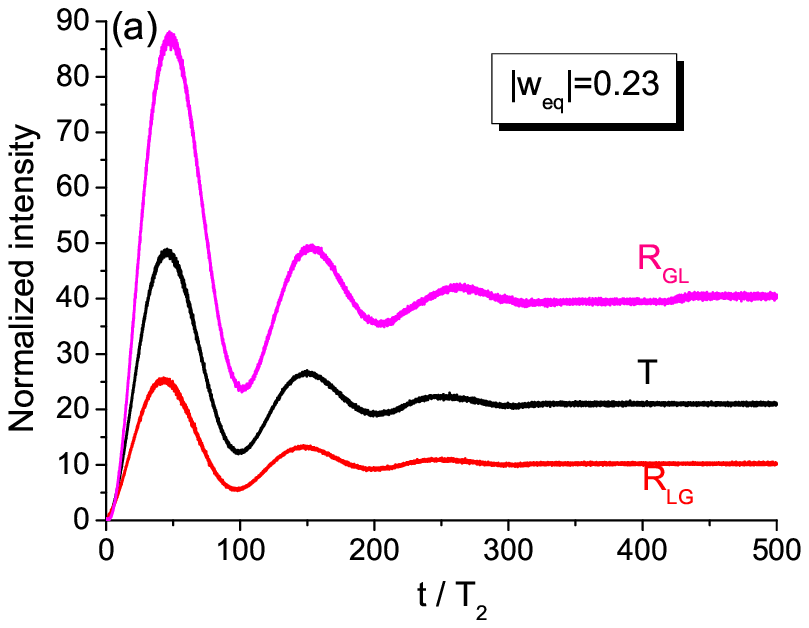}
\includegraphics[scale=1.0,clip=]{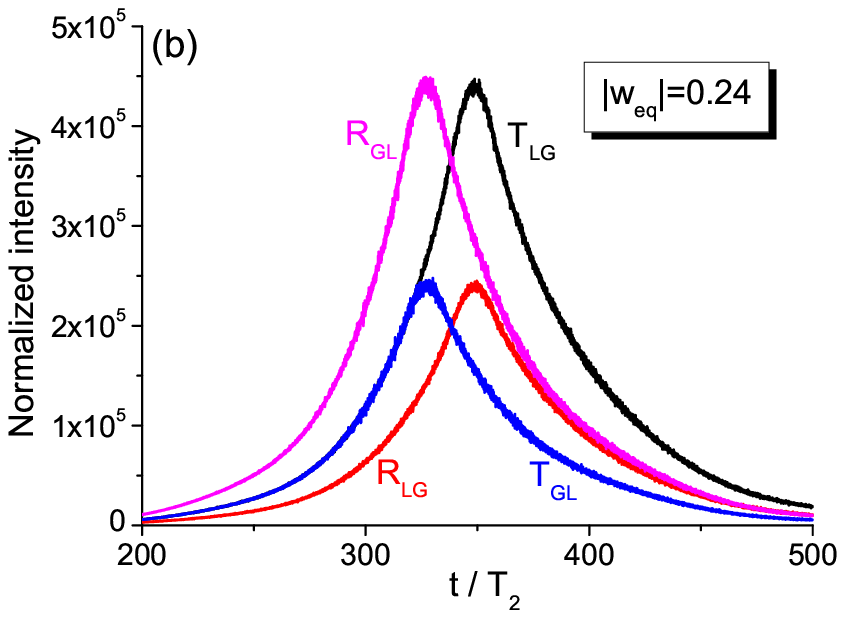}
\caption{The case of completely unexcited absorbing layers: Temporal
dynamics of the reflected (R) and transmitted (T) intensity for the
pumping parameter (a) $|w_{eq}|=0.23$ and (b) $|w_{eq}|=0.24$.}
\label{figS6}
\end{figure*}

The results of calculations with $\omega_{L,G}=10^{11}$ s$^{-1}$ and
$\omega_{L,L}=|w_{eq,G}| \omega_{L,G}$ for $|w_{eq}|=0.23$ and
$0.24$ are shown in Fig. \ref{figS6}. The other parameters are the
same as in the main text. It is seen that the profiles of
transmitted and reflected intensity are in good agreement with that
reported in Fig. 6. This proves that the cases of equal particles
density and completely unexcited absorbing layers are equivalent for
study of $\mathcal{PT}$ symmetry.


